\documentclass[aps,onecolumn]{revtex4}
\usepackage{amsmath}
\usepackage{amssymb,epsf}
\usepackage{xcolor}
\usepackage{soul}
\begin{document}

\title{New aspect of critical nonlinearly charged black hole}
\author{S. H. Hendi$^{1,2}$\footnote{email address: hendi@shirazu.ac.ir},
Z. S. Taghadomi$^{1}$ \footnote{email address:
z\_s\_taghadomi@yahoo.com} and C. Corda$^{2,3}$\footnote{email
address: cordac.galilei@gmail.com} } \affiliation{$^1$Physics
Department and Biruni Observatory, College of Sciences, Shiraz
University, Shiraz 71454, Iran\\
$^2$ Research Institute for Astronomy and Astrophysics of Maragha
(RIAAM), Maragha, Iran\\
$^{3}$ International Institute for Applicable Mathematics and
Information Sciences, Adarshnagar, Hyderabad 500063, India }

\begin{abstract}

The motion of a point charged particle moving in the background of
the critical power Maxwell charged AdS black holes, in a probe
approximation is studied. The extended phase space, where the
cosmological constant appears as a pressure, is regarded and the
effective potential is investigated. At last, the mass-to-charge
ratio and the large $q$ limit are studied.

\end{abstract}

%\keywords{wcwececwc ; wecwcecwc}

\maketitle

\section{Introduction}

Since the last decades, black hole (BH) solutions have been
considered the most interesting subjects of Einstein's general
theory relativity (GTR). Thermodynamic properties of these massive
objects have been renewed interest for many researchers since the
famous works of Hawking and Beckenstein {[}1-3{]}. In other words,
BH has been recognized as a thermodynamic system with the
Beckenstein-Hawking entropy and Hawking temperature
\cite{key-4,key-5}. Moreover, BH phase structure has shown
significant aspects of statistical mechanics. Once the space-time
metric is obtained, one can provide explicit equations for
thermodynamic quantities using the standard geometric
techniques. In this regard, the extended phase space
thermodynamics {[}6-9{]} supplies a pressure $p$ and volume $V$ to
supplement the vocabulary of the first law of thermodynamics
\cite{key-10}. Starting from an idea of Beckenstein \cite{key-11},
researchers in quantum gravity have today the intuitive, common
conviction that, in some respects, BHs are the fundamental bricks
of quantum gravity in the same way that atoms are the fundamental
bricks of quantum mechanics. This similarity suggests that the BH
mass should have a discrete spectrum \cite{key-11} and that BHs
result in highly excited states representing both the ``hydrogen
atom'' and the ``quasi-thermal emission'' in quantum gravity
\cite{key-12}. Some recent works have indeed shown that the
intuitive picture is more than a picture proposing interesting
approaches where BHs are really seen as being the ``gravitational
atoms'' \cite{key-12,key-13,key-14,key-15,key-16}.

In recent years, an interest in the subject of BH chemistry has
been growing. Within this context, one can consider a heat engine
which is a cycle in a pressure-volume space that extracts work
from the AdS BH \cite{key-17}. These kinds of engines are called
holographic heat engines, since one considers the negative
cosmological constant as a dynamical pressure. Another novel way
that reformulated the physics of charged BHs in the vicinity of
the critical point has been reported in {[}18-20{]}. In this way,
the physical quantities of the charged BHs are rescaled by the
electric charge. Following the method of Johnson
\cite{key-19,key-20}, one sees that the charged AdS BH behaves
like a Rindler space-time near the critical point if we regard the
double limit of near horizon and large charge, simultaneously.
Such behavior is quite interesting from the holographic gravity
point of view.

Standard Maxwell theory has been faced with some problems such as
infinite self-energy of point-like charges \cite{key-21,key-22},
vacuum polarization in quantum electrodynamics and low-energy
limit of heterotic string theory {[}23-31{]}. In addition, in
higher dimensions ($D>4$), the Maxwell action does not possess the
conformal symmetry \cite{key-32,key-33}. In general, when the
electromagnetic field is strong enough the linear electrodynamics
cannot support some theoretical evidence. In order to overcome
these problems, a class of non-linear electrodynamics (NLED) has
been introduced. This idea was first introduced by the pioneering
work of Born and Infeld to remove the mentioned infinite
self-energy \cite{key-21,key-22}. Moreover, based on Dirac
suggestion \cite{key-34}, that, due to the strong electromagnetic
field, one should consider the NLED near the point-like charges,
one may take into account the same behavior in the vicinity of
charged BHs \cite{key-35}. Another NLED model is called power
Maxwell invariant (PMI) theory. In this approach the Lagrangian
density is given by $(-F)^{s},$ being $s$ is an arbitrary rational
number. In such a model, one can adjust the NLED power in such a
way that the related action enjoys the conformal symmetry in
arbitrary dimensions. Recently, the NLED has been used in various
interesting astrophysical frameworks. In fact, the effects arising
from an NLED seem to become very important in super-strongly
magnetized compact objects, such as pulsars, and particular
neutron stars. Some examples include the so-called magnetars and
strange quark magnetars. In particular, NLED modifies in a
fundamental way the concept of gravitational redshift as compared
to the well-established method introduced by standard linear
electrodynamics \cite{key-36}. The analysis in \cite{key-36} has
shown that, unlike linear electrodynamics, where the gravitational
redshift is independent of any background magnetic field, when an
NLED is incorporated into the photon dynamics, an effective
gravitational redshift comes out, which happens to depend
decidedly on the magnetic field pervading the pulsar. A similar
result has also been obtained in \cite{key-37} for magnetars and
strange quark magnetars. The resulting gravitational redshift
tends to infinity as the magnetic field grows larger
\cite{key-36,key-37}, as opposed to the predictions of standard
linear electrodynamics. The particular importance of this work is
that the gravitational redshift of neutron stars is connected to
the mass\textendash radius relation of the object \cite{key-36}.
Thus, NLED effects turn out to be important regarding the
mass-radius relation, and one can also reasonably expect important
effects in the case of BH physics, where the mass-radius ratio is
even more important than for a neutron star. On the other hand, it
has been recently shown that NLED objects can remove singularities
too {[}38-45{]}. In particular, in \cite{key-38} the density
singularity of a BH has been removed in a way that, at the end,
fixed the radius of the star, the final density depends only on
the star mass and on a \emph{quintessential density term} which
arises from the Heisenberg-Euler NLED \cite{key-38}
\begin{equation}
\mathcal{L}_{m}\equiv-\frac{1}{4}F+c_{1}F^{2}+c_{2}G^{2},\label{eq: 3}
\end{equation}
where
$G=\frac{1}{2}\eta_{\alpha\beta\mu\nu}F^{\alpha\beta}F^{\mu\nu}$,
$F\equiv F_{\mu\nu}F^{\mu\nu}$ is the electromagnetic scalar and
$c_{1}$ and $c_{2}$ are constants, see \cite{key-38} for details.
In this work, we are going to explore some thermodynamical points
of BHs in the PMI gravity with considering the extended phase
space. PMI theory is one of the interesting NLED models. As one
knows, due to the existence of a singularity in the linear Maxwell
electrodynamics, one motivates to study the NLED effects on the BH
thermodynamic properties in the extended phase space. The main
focus of the paper is investigating the new effects of PMI theory
on the BH. Some of the interesting properties of BH thermodynamics
in Einstein-PMI theory have been studied before {[}32, 33,
46-52{]}. In addition, the effects of the PMI NLED source on the
strongly coupled dual gauge theory have been studied in the
context of AdS/CFT correspondence \cite{key-53,key-54}.

Here, we are focussing on a specific point in the parameter space
where its properties lie at a critical point. It is motivated by
the recent approach of Johnson \cite{key-20}. In that work, a new
limit where the BH charge is taking to large values has been
introduced. In this regime, a specific heat engine working close
to the Carnot efficiency \cite{key-20}.

We also take the chance to stress that there are also motivations
beyond gravitational theory which suggest that critical BHs
deserve to be studied. In fact, the thermodynamics arising from
the approach in this work define interesting thermodynamic
equations that could be, in principle, used in other fields of
physics, for example in statistical mechanics
\cite{key-55,key-56}.

As it follows, after introducing the space-time metric and the
thermodynamic quantities, in brief, we look closely at the
critical point. Then, we obtain some interesting new results at
the critical point and discuss them.

\section{Einstein-PMI gravity in anti-de Sitter space-time}

The bulk action of Einstein-PMI gravity in a $D-$dimensional
anti-de Sitter space-time has the following form {[}32, 33,
46-52{]}
\begin{equation}
I=-\frac{1}{16\pi}\int\!d^{D}x\sqrt{-g}\left(R-2\Lambda+(-F)^{s}\right).\label{eq:action}
\end{equation}
where $F=F_{\mu\nu}F^{\mu\nu}$, $s$ is the nonlinearity parameter
and $\Lambda=-(D-1)(D-2)/2l^{2}$ is the negative cosmological
constant with a length scale $l$ ( we work with $G=c=1$).

Let us consider a spherically symmetric space-time as
\begin{eqnarray}
ds^{2} & = & -Y(r)dt^{2}+\frac{dr^{2}}{Y(r)}+r^{2}d\Omega_{D-2}^{2},\label{eq:ds}
\end{eqnarray}
where $d\Omega_{D-2}$ represents the standard line element of a $D-2$
sphere ($S^{D-2}$). If one considers the field equations arising
from the variation of the bulk action with the metric ($\ref{eq:ds}$),
one can show that the metric function $Y(r)$, the gauge potential
one-form $A$ and the electromagnetic field two-form $F$ are given
by {[}32, 33, 46-52{]}:
\begin{eqnarray}
Y(r) & = & 1-\frac{m}{r^{D-3}}+\frac{r^{2}}{l^{2}}+\frac{(2s-1)^{2}
\left(\frac{(D-2)(2s-D+1)^{2}q^{2}}{(D-3)(2s-1)^{2}}\right) ^{s}}
{(D-2)(D-1-2s)r^{2\left(\frac{Ds-4s+1}{(2s-1)}\right)}},\\
A & = & -\sqrt{\frac{D-2}{2(D-3)}}qr^{\frac{2s-D+1}{2s-1}}~dt,\\
F & = & dA.
\end{eqnarray}
The nonlinearity parameter of the source is restricted to
$s>\frac{1}{2}$ {[}32, 33, 46-52{]}, and the parameters $m$ and
$q$ are, respectively, related to the BH ADM mass, $M$, and the BH
electric charge, $Q,$ as
\begin{equation}
M=\frac{\omega_{D-2}~(D-2)}{16\pi}m,\label{eq:mass}
\end{equation}

\begin{equation}
Q=\frac{\omega_{D-2}~\sqrt{2}(2s-1)s}{8\pi}\left(\frac{D-2}{D-3}\right)
^{s-\frac{1}{2}}\left(\frac{(D-1-2s)q}{2s-1}\right)^{2s-1},\label{eq:charge}
\end{equation}
where $\omega_{D-2}$ is the volume of a unit $(D-2)$ sphere, given
by
\begin{equation}
\omega_{D-2}=\frac{2\pi^{\frac{D-1}{2}}}{\Gamma(\frac{D-1}{2})}.\label{eq:omega}
\end{equation}
The BH event horizon radius can be calculated numerically by finding
the largest real positive root of the metric function $Y(r=r_{+})=0$.
Using the surface gravity ($\kappa$) conception, one can obtain the
Hawking temperature
\begin{equation}
\begin{array}{c}
T=\frac{\kappa}{2\pi}=\frac{Y'(r_{+})}{4\pi}=\\
\\
\frac{D-3}{4\pi r_{+}}\left[1+\frac{D-1}{D-3}
\frac{r_{+}^{2}}{l^{2}}-\frac{(2s-1)\left(\frac{(D-2)(2s-D+1)^{2}
q^{2}}{(D-3)(2s-1)^{2}}\right)^{s}}{(D-2)(D-3)r_{+}^{2\frac{Ds-4s+1}{2s-1}}}\right].
\end{array}\label{eq: temperature}
\end{equation}
As it was considered in {[}6-9{]}, one can interpret $\Lambda$ as
a thermodynamic pressure $P$:
\begin{equation}
P=\frac{-\Lambda}{8\pi}=\frac{(D-1)(D-2)}{16\pi l^{2}}\ ,\label{eq:pressure}
\end{equation}
where its corresponding conjugate quantity is the thermodynamic
volume which is \cite{key-57}
\begin{equation}
V=\frac{\omega_{D-2}r_{+}^{D-1}}{D-1}.\label{eq:volume}
\end{equation}
Using Eqs. ($\ref{eq: temperature}$) and ($\ref{eq:pressure}$) for
a fixed charge $Q$, one obtains the following equation of state
\begin{equation}
\begin{array}{c}
P(T,r_{+})=\frac{D-2}{4r_{+}}T-\frac{(D-2)(D-3)}{16\pi r_{+}^{2}}+\\
\\
\frac{1}{16\pi}\frac{(2s-1)\left(\frac{(D-2)(2s-D+1)^{2}q^{2}}
{(D-3)(2s-1)^{2}}\right)^{s}}{r_{+}^{2s\frac{D-2}{2s-1}}}.
\end{array}\label{eq:pvstate}
\end{equation}
In addition, vanishing the metric function at the event horizon ($Y(r_{+})=0$)
can be rearranged in order to obtain an equation for the geometric
mass, $m$, which yields the total mass $M$ via Eq. ($\ref{eq:mass}$)
as
\begin{equation}
\begin{array}{c}
m=r_{+}^{D-3}+\frac{16\pi pr_{+}^{D-1}}{(D-1)(D-2)}+\\
\\
\frac{4s^{2}-4s+1}{r_{+}^{\frac{D-2s-1}{2s-1}}(D^{2}-2Ds-3D+4s+2)}
(\frac{(D-2)(D-2s-1)^{2}q^{2}}{(D-3)(2s-1)^{2}})^{s}.
\end{array}\label{eq:ml}
\end{equation}
One can use Eq. (\ref{eq:ml}) to obtain the volume \cite{key-10}.
The main focus here will be the critical point, a point of inflection
on the critical isotherm curve of the $P-V$ diagram. In fact,
\begin{equation}
\left(\frac{\partial P}{\partial V}\right)_{T}=0,\left(\frac{\partial^{2}P}{\partial V^{2}}\right)_{T}=0.
\end{equation}
Being $r_{+}$ related to the specific volume {[}6-9{]}, the recent
derivatives can also be taken with respect to the event horizon,
$r_{+}$. As a result, one obtains the critical point as
\cite{key-18}
\begin{equation}
\begin{array}{c}
r_{cr}= \frac{D-2}{4}\left(\frac{16^{\frac{Ds-4s+1}{2s-1}}
(\frac{(D-2)(D-2s-1)^{2}}{(D-3)(2s-1)^{2}})^{s}(D-2)^
{\frac{-2(Ds-4s+1)}{2s-1}}(2Ds-6s+1)sq^{2s}}{(D-3)(2s-1)}\right)
^{\frac{2s-1}{2(Ds-4s+1)}},
\end{array}\label{r_cr}
\end{equation}

\begin{equation}
T_{cr}=\frac{4(D-3)(Ds-4s+1)\left(\frac{ks(D-2)^{2}
(2Ds-6s+1)q^{2s}}{16(D-3)(2s-1)^{2}}\right)^{\frac{1-2s}
{2(Ds-4s+1)}}}{\pi(D-2)(2Ds-6s+1)},\label{eq: T_cr}
\end{equation}

\begin{equation}
P_{cr}=\frac{(D-3)(Ds-4s+1)}{\pi s(D-2)^{2}\left(\frac{ks(D-2)^{2}
(2Ds-6s+1)q^{2s}}{16(D-3)(2s-1)^{2}}\right)^{\frac{2s-1}{Ds-4s+1}}},\label{P_cr}
\end{equation}
 where
\[
k=\frac{16^{\frac{s(D-2)}{2s-1}}(2s-1)\left(\frac{(D-2)(2s-D+1)^{2}}
{(D-3)(2s-1)^{2}}\right)^{s}}{(D-2)^{\frac{2s(D-2)}{2s-1}}}.
\]
The four dimensional linear case ($D=4$ and $s=1$) is special. It
is easy to show that
\begin{equation}
r_{{\rm cr}}=\sqrt{6}q,\quad T_{{\rm cr}}=\frac{1}{3\sqrt{6}\pi q},
\quad P_{{\rm cr}}=\frac{1}{96\pi q^{2}}.\label{eq:critical4}
\end{equation}
As we mentioned before, the cosmological constant acts as a
thermodynamic variable in different theories of gravity. The
cosmological constant exists in various actions of gravity and
also one can trace it in the solution of such theories. These can
lead to use the BH as a heat engine. It was shown that the
efficiency of such heat engines may asymptote to the Carnot
efficiency in the limit of large $q$ which is different from the
ideal gas or correspondingly the high temperature limit
\cite{key-20}.

Taking into account the above statements, one is motivated by the
recent results mentioned to look closely at the critical region
for the large limit of $q$. One focuses on the charged black hole
in the presence of PMI NLED, since the PMI field is significantly
richer than the Maxwell field. In addition, in the specific limit
of the nonlinearity parameter, $s=1$, our theory reduces to the
Maxwell field, as expected. Using Eqs. ($\ref{eq:pressure}$),
($\ref{eq:pvstate}$), ($\ref{eq:ml}$) and
(\ref{r_cr})-(\ref{P_cr}), the critical BH mass parameter,
$m_{{\rm cr}}$, and the critical cosmological length scale,
$l_{{\rm cr}}$, are computed as
\begin{equation}
\begin{array}{c}
l_{{\rm cr}}^{2}=\frac{\left(\frac{16^{\frac{Ds-4s+1}{2s-1}}
(\frac{(D-2)(D-2s-1)^{2}}{(D-3)(2s-1)^{2}})^{s}(D-2)^{\frac{-2(Ds-4s+1)}
{2s-1}}s(2Ds-6s+1)q^{2s}}{(D-3)(2s-1)}\right)^{\frac{2s-1}{Ds-4s+1}}}
{16(sD^{2}-7Ds+D+12s-3)},\\
\\
m_{cr}=m(r_{+}=r_{cr}).
\end{array}\label{eq: parameter}
\end{equation}
Considering the special case $D=4$ and $s=1$, one obtains $l_{{\rm cr}}^{2}=36q^{2}$
and $M_{{\rm cr}}=m_{{\rm cr}}/2=4/\sqrt{6}q$.

Now, one is in a position to study the critical BH region by inserting
the critical values. Using the obtained relations, one rewrites the
metric function for the critical BH in the following form
\begin{equation}
Y_{{\rm cr}}(r)=1-\frac{m_{{\rm cr}}}{r^{D-3}}+\frac{r^{2}}{l_{{\rm cr}}^{2}}+
\frac{(2s-1)^{2}\left(\frac{(D-2)(2s-D+1)^{2}q^{2}}{(D-3)(2s-1)^{2}}\right)^{s}}
{(D-2)(D-1-2s)r^{2\left(\frac{Ds-4s+1}{(2s-1)}\right)}}.\label{metric}
\end{equation}
Now, one can study the motion of a point particle of mass $\mu$
and charge $e$ moving in the background of the obtained critical,
nonlinearly charged BH in a probe approximation.

In order to obtain effective potential, we apply Hamilton-Jacobi
formalism as a standard technique \cite{key-58}. The
Hamilton-Jacobi equation for a geometry described by the metric
$g_{\mu \nu}$ is
\begin{equation}
g^{\mu\nu}(\frac{\partial S}{\partial x^{\mu}}+
qA_{\mu})(\frac{\partial S}{\partial
x^{\nu}}+qA_{\nu})+\mu^2=0,\label{eq:hamiltonjacobi}
\end{equation}
where $A_{\mu}$ is the vector potential and we consider the
following ansatz for $S$
\begin{equation}
S=Et+S_1(r)+S_2(\theta)+J\phi,\label{eq:ansatz}
\end{equation}

Since we have considered static charged black hole, the only
non-vanishing component of $A_{\mu}$ is
\begin{equation}
A_{t}=-q\sqrt{\frac{D-2}{2(D-3)}}
r^{\frac{2s-D+1}{2s-1}}.\label{eq:amu}
\end{equation}

Using the metric introduced in Eq. (\ref{eq:ds}) and the ansatz
for $S$, Eq. (\ref{eq:hamiltonjacobi}) can be rewritten as
\begin{equation}
-\frac{r^2}{Y(r)}\left[E-eq\sqrt{\frac{D-2}{2(D-3)}}
r^{\frac{2s-D+1}{2s-1}}\right]^2+r^2Y(r)\left(\frac{\partial
S_1}{\partial
r}\right)^2+L^2+r^2\mu^2=0,\label{eq:hamiltonjacobi2}
\end{equation}
where $L$ is the angular momentum of the particle which is defined
as $L=\sqrt{(\frac{\partial S_2}{\partial
\theta})^2+J^2csc(\theta)^2}$. From Eq. (\ref{eq:hamiltonjacobi}),
$S_1(r)$ reads as follow
\begin{equation}
S_1(r)=\pm
\int{\frac{dr}{Y(r)}\sqrt{[E-q\;e\;\sqrt{\frac{D-2}{2(D-3)}}
r^{\frac{2s-D+1}{2s-1}}]^2-Y(r)(\frac{L^2}{r^2}+\mu^2)}}.\label{eq:s1r}
\end{equation}

Using the fact that $\frac{\partial t}{\partial
r}=\frac{\partial}{\partial r} (\frac{\partial S}{\partial
E})=\frac{\partial}{\partial E}(\frac{\partial S}{\partial
r})=\frac{\partial}{\partial E}(\frac{\partial S_1(r)}{\partial
r})$, one finds
\begin{equation}
\frac{dr}{dt}=\pm \frac{Y(r)}{E-
q\;e\;\sqrt{\frac{D-2}{2(D-3)}}r^{\frac{2s-D+1}{2s-1}}}
\sqrt{[E-q\;e\;\sqrt{\frac{D-2}{2(D-3)}}r^{\frac{2s-D+1}
{2s-1}}]^2-Y(r)(\frac{L^2}{r^2}+\mu^2)}.\label{eq:drdt}
\end{equation}

The condition of turning point $\frac{dr}{dt}=0$ allows us to
define the following effective potential
\begin{equation}
V_{{\rm
eff}}=\sqrt{\frac{D-2}{2D-6}}q\;e\;r^{\frac{2s-D+1}{2s-1}}+\sqrt{Y_{{\rm
cr}}(r)}
\sqrt{\mu^{2}+\frac{L^{2}}{r^{2}}}.\label{eq:effectivepotential}
\end{equation}

In general, this potential may have a local minimum at some
$r_{{\rm min}}>r_{{\rm cr}}$. This depends on the values of $e$,
$\mu$ and $L$. Now, let us look for the physical minimums for the
case of $L=0$ and $r_{{\rm min}}>r_{{\rm cr}}$. In this case the
probe has specific critical mass-to-charge ratio ($\mu/e$) for
different variables like dimensions, $D$, nonlinearity parameter,
$s$, and BH electric charge, $q$.

The numeric results for the case of different dimensions, $D$,
with $q=10$ and $L=0$ are shown in Table \ref{tab11}. In addition,
we complete the variation of $s$ and $q$ in Table \ref{tab22}.

\begin{figure}[tbp]
$
\begin{array}{c}
\epsfxsize=6.5cm \epsffile{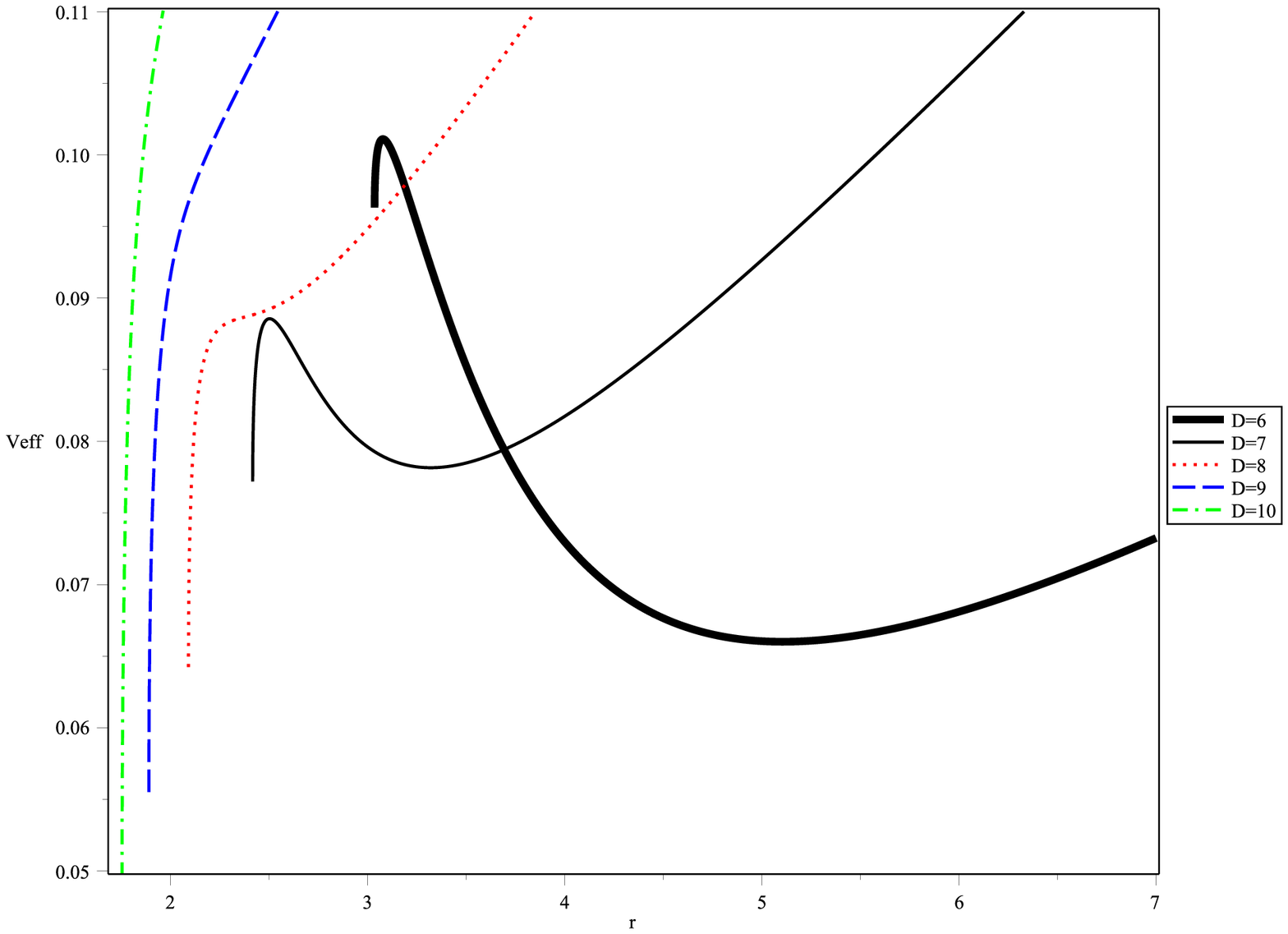} \\
\epsfxsize=6.5cm \epsffile{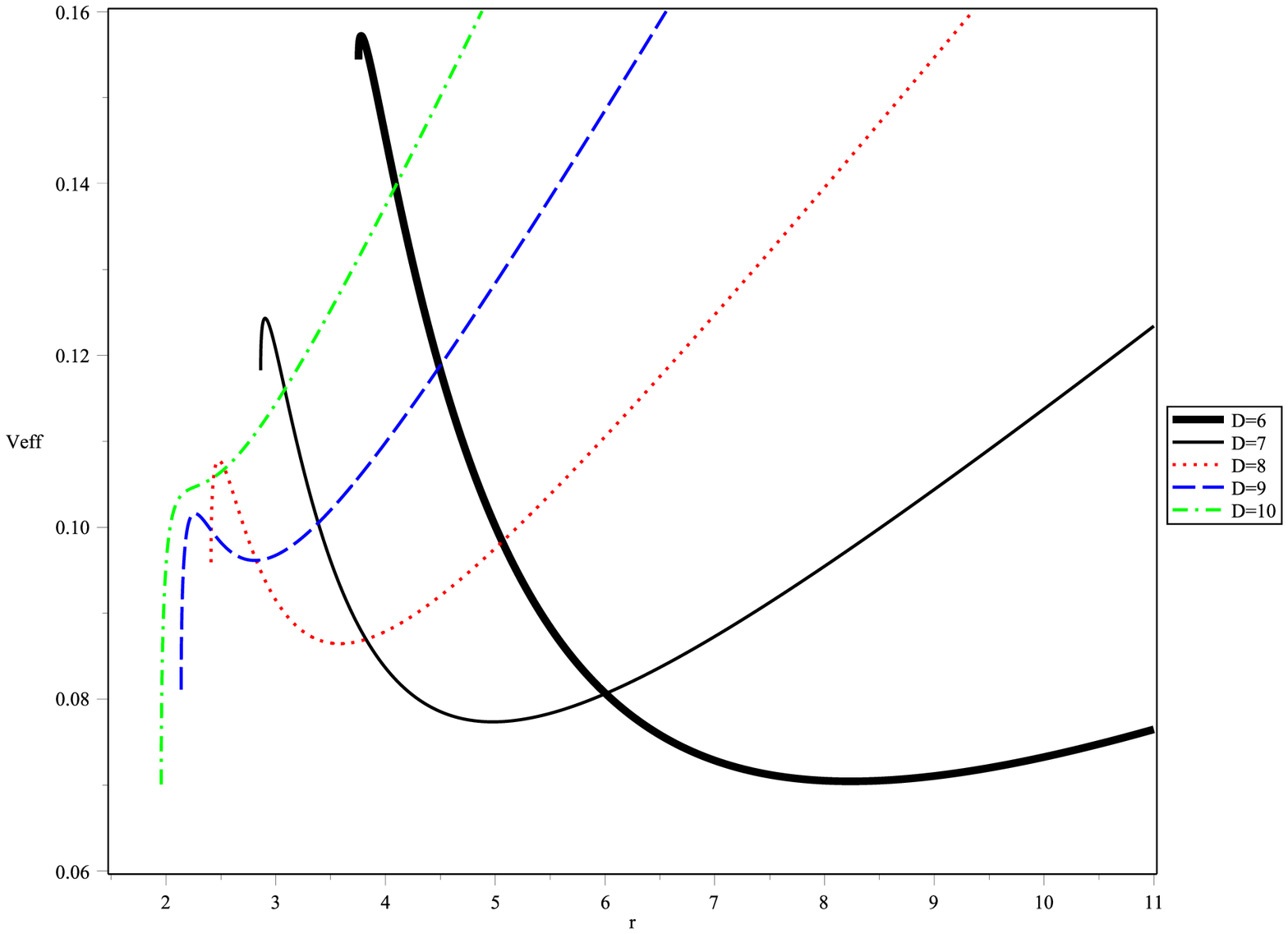} \\
\epsfxsize=6.5cm \epsffile{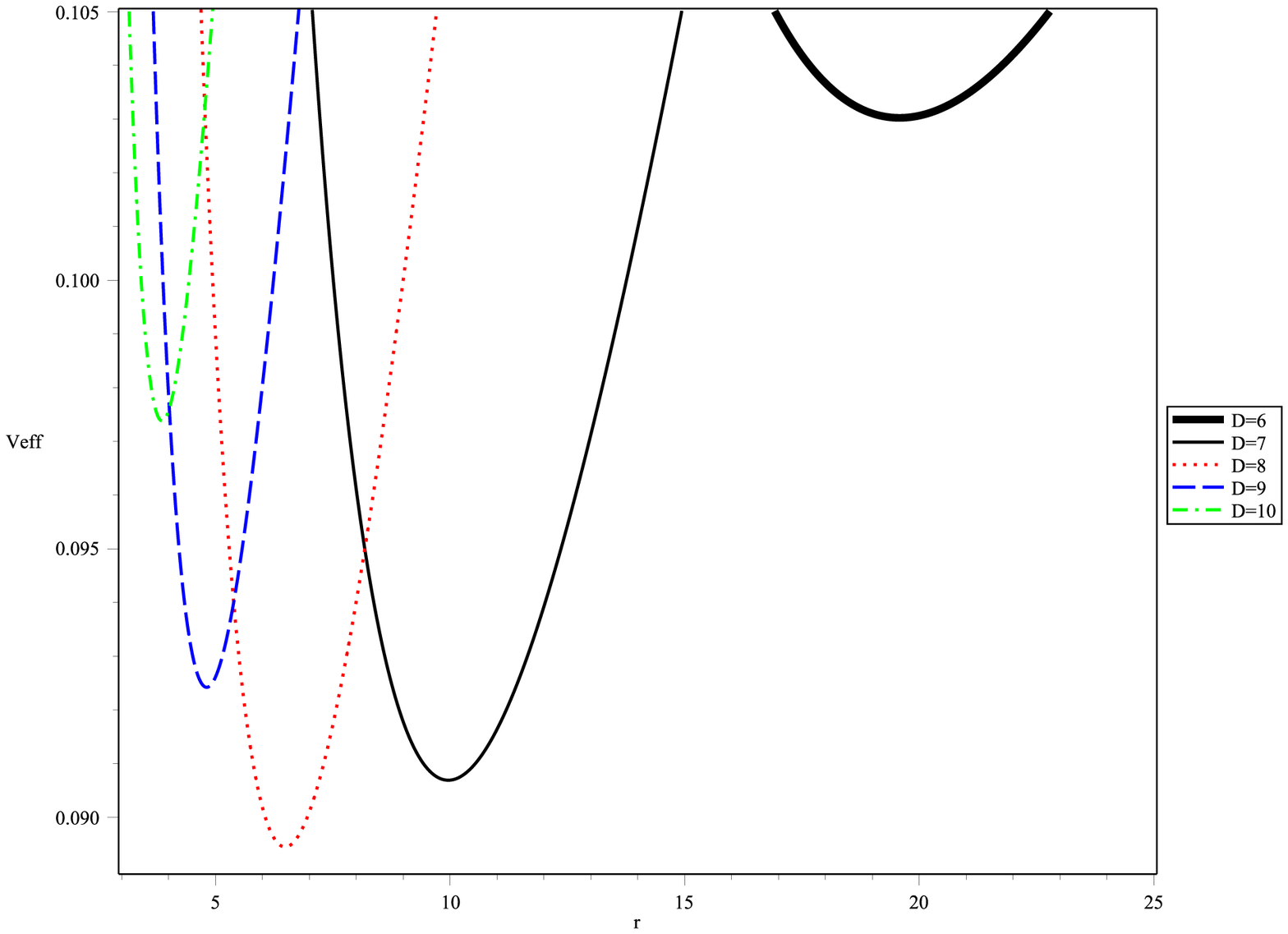}
\end{array}
$ \caption{$V_{eff}(r)$ versus $r$ for $L = 0$ and $q=10$, and
$D=6$ black (bold line), $D=7$ black (solid line), $D=8$ red
(dotted line), $D=9$ blue (dashed line) and $D=10$ green
(dash-dotted line) with s=0.9 (\textbf{up panel}), $s = 1$
\textbf{(Middle panel)} and $s=1.2$ \textbf{(down panel)}.}
\label{various-D}
\end{figure}

Based on Fig. \ref{various-D}, If we consider $s \leq 1$, there is
a $D_{0}$ in which $V_{eff}$ is an increasing function of $r$ for
$D>D_{0}$ while for $D<D_{0}$ it has a local maximum and an
absolute minimum. By increasing dimensionality to reach $D_{0}$,
one finds that the increasing minimum takes place in smaller $r$.
Regarding $s>1$, it is shown that by increasing dimensions,
although values of $r_{min}$ is a decreasing function, $V_{min}$
is a decreasing function of $r$ for $D<D{c}$ and after $D{c}$, it
is an increasing function of $r$. The lowest amount of $V_{min}$
takes place for $D_{c}$, in which $D_{c}$ depends on the
nonlinearity parameter, $s$.

\begin{figure}[tbp]
$
\begin{array}{c}
\epsfxsize=6.5cm \epsffile{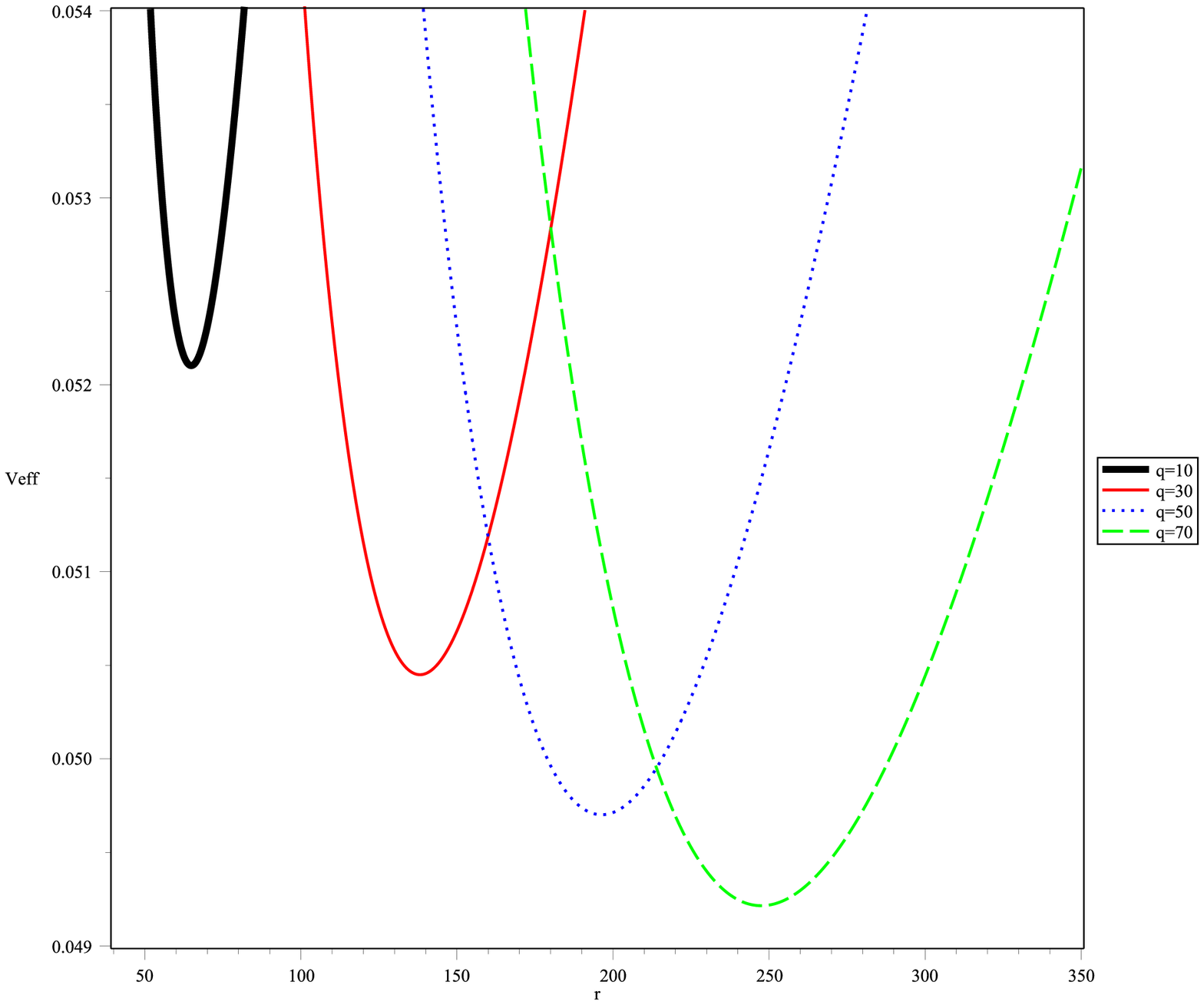} \\
\epsfxsize=6.5cm \epsffile{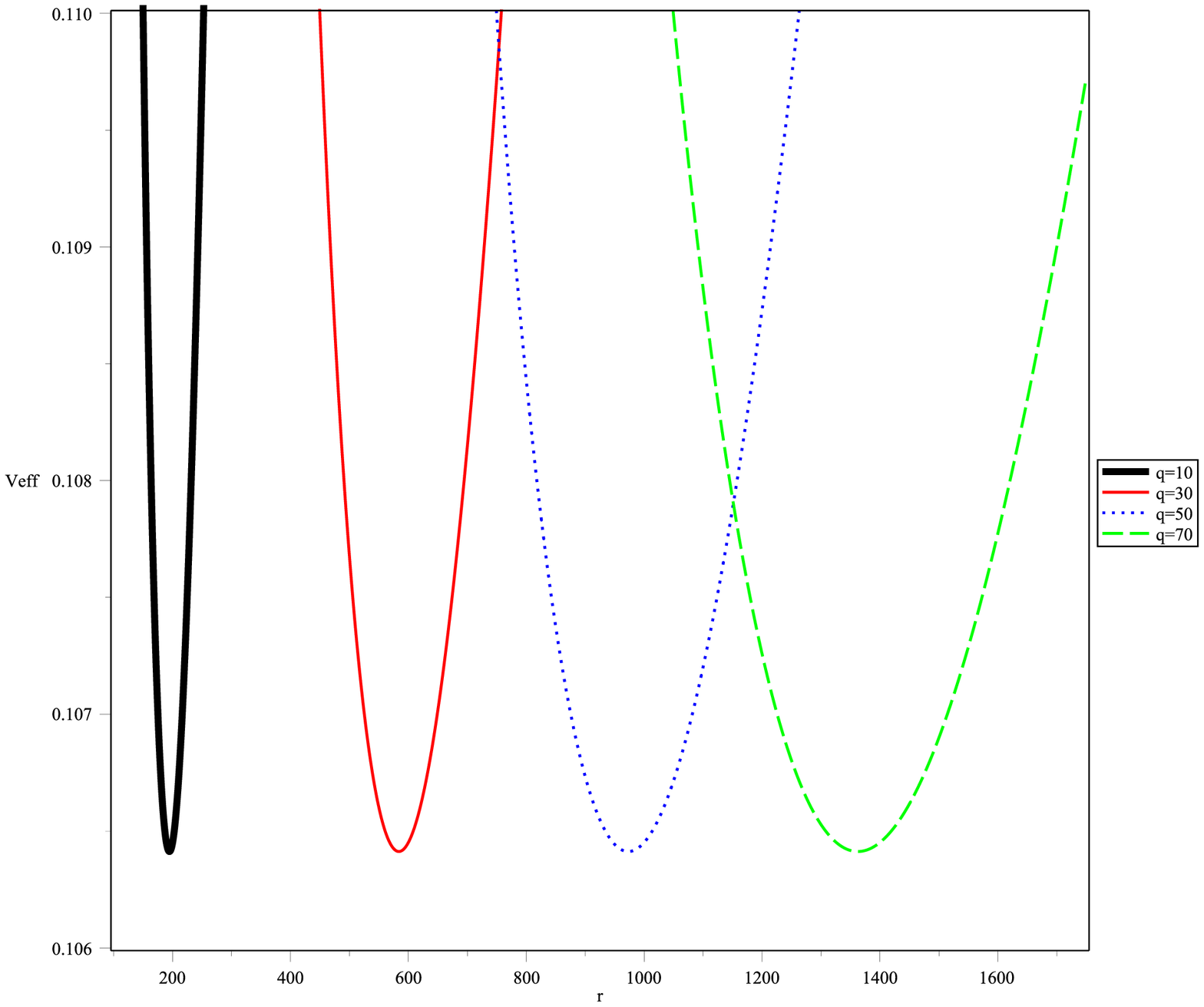} \\
\epsfxsize=6.5cm \epsffile{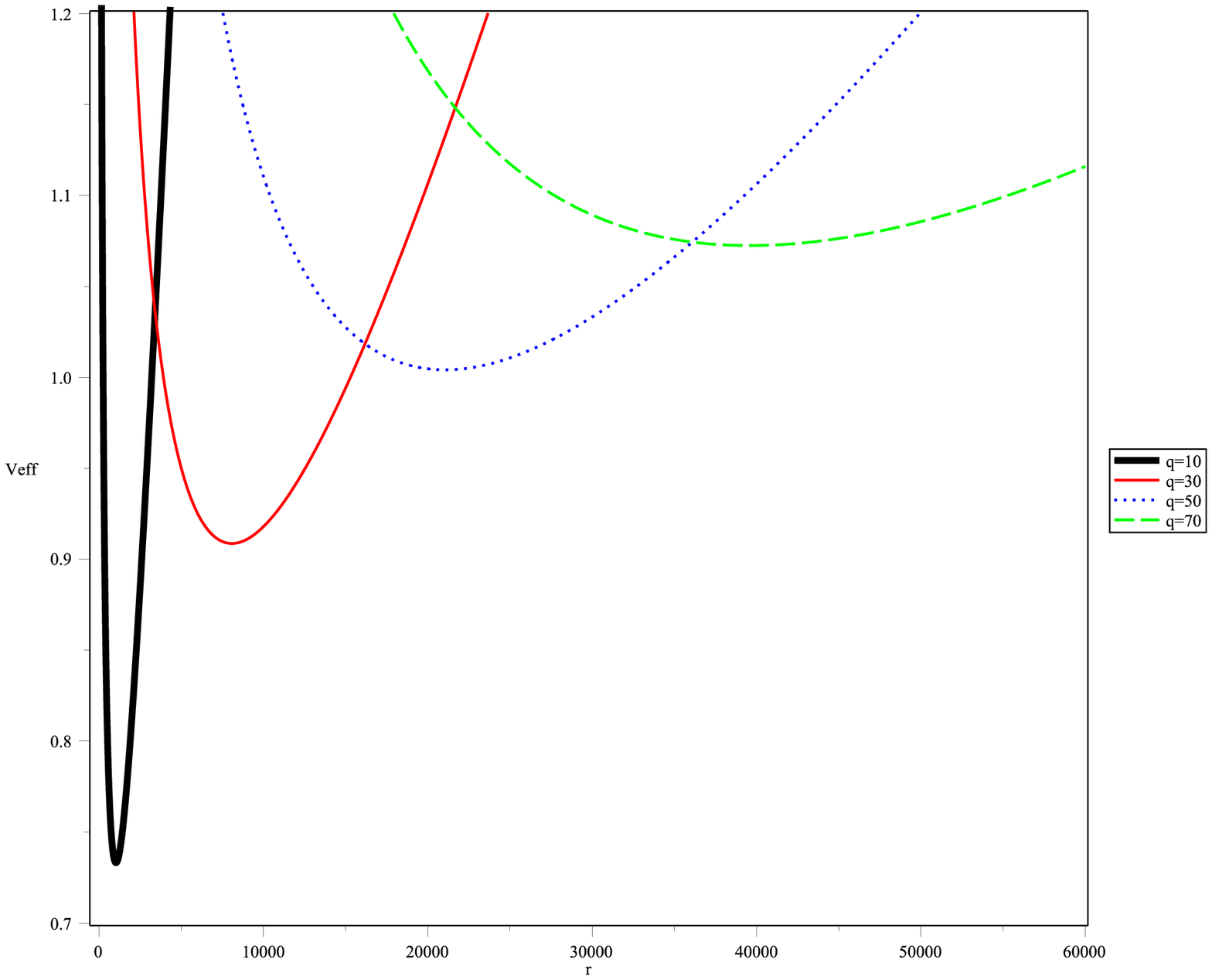}
\end{array}
$ \caption{$V_{eff}(r)$ versus $r$ for $L = 0$ and $D=4$, and
$q=10$ black (bold line), $q=30$ red (solid line), $q=50$ blue
(dotted line) and $q=70$ green (dashed line) with s=0.9
(\textbf{up panel}), $s = 1$ \textbf{(Middle panel)} and $s=1.2$
\textbf{(down panel)}.} \label{various-q}
\end{figure}

Having a look to all panels of Fig. \ref{various-q}, one can find
an interesting result. It is clear that by increasing the electric
charge in $s<1$ case, $V_{min}$ is a decreasing function of $q$,
while it is an increasing function of the
 electric charge for $s>1$. The middle panel of
Fig. \ref{various-q} indicates that for linear case (Maxwell
field), $V_{min}$ does not depend on $q$.

%%%%%%%%%%%%%%%%%%%%%%%%%%%%%%%%%%%%%%%%%%%%%%%%%%%%%%%%%%%%%%%%%%%%%%%%%%%%%%%%%%%%%%%%%%%%
%%%%%%%%%%%%%%%%%%%%%%%%%%%%%%%%%%%%%%%%%%%%%%%%%%%%%%%%%%%%%%%%%%%%%%%%%%%%%%%%%%%%%%%%%%%%%%%%%%%%%%%%%%%%%%%%%%%%%%%%%%%%%%%%

\begin{table}[tbp]
    \centering
    \begin{tabular}[t]{|c|c|c|c|c|c|}
        \toprule
        \hline\hline $D$ & $m_{cr}$ & $\frac{M_{cr}}{Q}$ & $l_{cr}$ & $r_{min}$ & $V_{min}$\\
        \hline\hline $6$ & $41.64$ & $3.81$ & $4.44$ & $5.11$ & $0.066$ \\
        \hline
        $7$ & $53.34$ & $5.04$ & $3.26$ & $3.32$ & $0.078$ \\
        \hline$8$ & $64.83$ & $6.29$ & $2.68$ & $-$ & $-$ \\
        \hline $9$ & $76.07$ & $7.57$ & $2.34$ & $-$ & $-$  \\
         \hline $10$ & $87.08$ & $8.86$ & $2.11$ & $-$ & $-$ \\ \hline
    \end{tabular}\hfill%
    \begin{tabular}[t]{|c|c|c|c|c|c|}
        \toprule
        \hline\hline $D$ & $m_{cr}$ & $\frac{M_{cr}}{Q}$ & $l_{cr}$ & $r_{min}$ & $V_{min}$ \\
        \hline\hline $6$ & $78.62$ & $3.21$ & $5.60$ & $8.24$ & $0.070$ \\
        \hline
        $7$ & $104.35$ & $4.12$ & $3.92$ & $4.98$ & $0.077$ \\
        \hline$8$ & $130.83$ & $5.07$ & $3.12$ & $3.59$ & $0.086$ \\
        \hline $9$ & $157.77$ & $6.02$ & $2.67$ & $2.81$ & $0.096$  \\
        \hline $10$ & $185.01$ & $6.02$ & $2.37$ & $-$ & $-$ \\  \hline
    \end{tabular}\hfill%
    \begin{tabular}[t]{|c|c|c|c|c|c|}
    \toprule
    \hline\hline $D$ & $m_{cr}$ & $\frac{M_{cr}}{Q}$ & $l_{cr}$ & $r_{min}$ & $V_{min}$ \\
    \hline\hline $6$ & $239.61$ & $2.76$ & $8.31$ & $19.58$ & $0.103$ \\
    \hline
    $7$ & $347.22$ & $3.32$ & $5.40$ & $9.98$ & $0.091$ \\
    \hline $8$ & $467.67$ & $3.91$ & $4.10$ & $6.48$ & $0.089$ \\
    \hline $9$ & $599.11$ & $4.53$ & $3.38$ & $4.81$ & $0.092$ \\
    \hline $10$ & $740.24$ & $5.15$ & $2.93$ & $3.86$ & $0.097$ \\ \hline
\end{tabular}\hfill
    \caption{The numeric results of $m_{cr}$, the critical mass\--to\--charge ratio for
        the black hole($\frac{M_{cr}}{Q})$ and $l_{cr}$ for $L=0$, $q=10$ and different dimensions. The left, middle and right tables have nonlinearity parameters $s=0.9$, $s=1$ and $s=1.2$ respectively.}\label{tab11}
\end{table}

%%%%%%%%%%%%%%%%%%%%%%%%%%%%%%%%%%%%%%%%%%%%%%%%%%%%%%%%%%%%%%%%%%%%%%

\begin{table}[tbp]
    \centering
    \begin{tabular}[t]{|c|c|c|c|c|c|}
        \toprule
        \hline\hline $q$ & $m_{cr}$ & $\frac{M_{cr}}{Q}$ & $l_{cr}$ & $r_{min}$ & $V_{min}$\\
        \hline\hline $10$ & $18.12$ & $1.55$ & $33.16$ & $64.92$ & $0.052$ \\
        \hline
        $30$ & $39.97$ & $1.42$ & $73.15$ & $138.10$ & $0.050$ \\
        \hline$50$ & $57.74$ & $1.36$ & $105.67$ & $196.40$ & $0.0497$ \\
        \hline $70$ & $73.57$ & $1.32$ & $134.63$ & $247.65$ & $0.049$  \\
         \hline
    \end{tabular}\hfill%
    \begin{tabular}[t]{|c|c|c|c|c|c|}
        \toprule
        \hline\hline $q$ & $m_{cr}$ & $\frac{M_{cr}}{Q}$ & $l_{cr}$ & $r_{min}$ & $V_{min}$ \\
        \hline\hline $10.0$ & $32.66$ & $1.63$ & $60.0$ & $194.63$ & $0.106$ \\
        \hline
        $30.0$ & $97.98$ & $1.63$ & $180.0$ & $778.51$ & $0.106$ \\
        \hline$50$ & $163.30$ & $1.63$ & $300.0$ & $973.14$ & $0.106$ \\
        \hline $70.0$ & $228.62$ & $1.63$ & $420.0$ & $1362.39$ & $0.106$  \\
         \hline
    \end{tabular}\hfill%
    \begin{tabular}[t]{|c|c|c|c|c|c|}
        \toprule
        \hline\hline $q$ & $m_{cr}$ & $\frac{M_{cr}}{Q}$ & $l_{cr}$ & $r_{min}$ & $V_{min}$ \\
        \hline\hline $10$ & $74.15$ & $2.50$ & $117.08$ & $1030.26$ & $0.733$ \\
        \hline
        $30$ & $469.55$ & $3.41$ & $741.42$ & $8081.10$ & $0.909$ \\
        \hline $50$ & $1107.61$ & $3.93$ & $1748.93$ & $21060.19$ & $1.004$ \\
        \hline $70$ & $1949.32$ & $4.32$ & $3077.99$ & $39581.09$ & $1.072$ \\
         \hline
    \end{tabular}\hfill
    \caption{The numeric results of $m_{cr}$, the critical mass\--to\--charge ratio for
        the black hole($\frac{M_{cr}}{Q})$ and $l_{cr}$ for $L=0$, $D=4$ and different $q$. The left, middle and right tables have nonlinearity parameters, $s=0.9$, $s=1$ and $s=1.2$ respectively.}\label{tab22}
\end{table}

%%%%%%%%%%%%%%%%%%%%%%%%%%%%%%%%%%%%%%%%%%%

According to the Table \ref{tab11} and Table \ref{tab22}, at fixed
$q$ or dimension, $D$, by increasing the nonlinearity parameter,
$s$, the $r_{min}$ increases. In addition, as it is shown in Fig.
\ref{various-D} and Fig. \ref{various-q} the range that probe
particle is oscillating within is expanding by increasing $D$ or
$q$. Interestingly, when the nonlinearity parameter, $s$,
increases, the minimum of effective potential increases too. This
means that for different nonlinearity parameters the probe
particle has different energies in its equilibrium state. Such
energies will increase by increasing the nonlinearity parameter.

Furthermore, we investigate the case for different dimensions in
the linear case, $s=1$, and $q=10$. The results are shown in the
middle table of Table \ref{tab11} and also in the middle panel of
Fig. \ref{various-D}. Such results indicate that, as one seeks
minimum for different dimensions, the $r_{min}$ decreases with
increasing dimensions. In addition, as $D$ increases, the minimum
energy of the probe particle in its equilibrium state rises.

\section{Effective potential in the limit of large $q$}

\begin{figure}[tbp]
$%
\begin{array}{cc}
\epsfxsize=6.5cm \epsffile{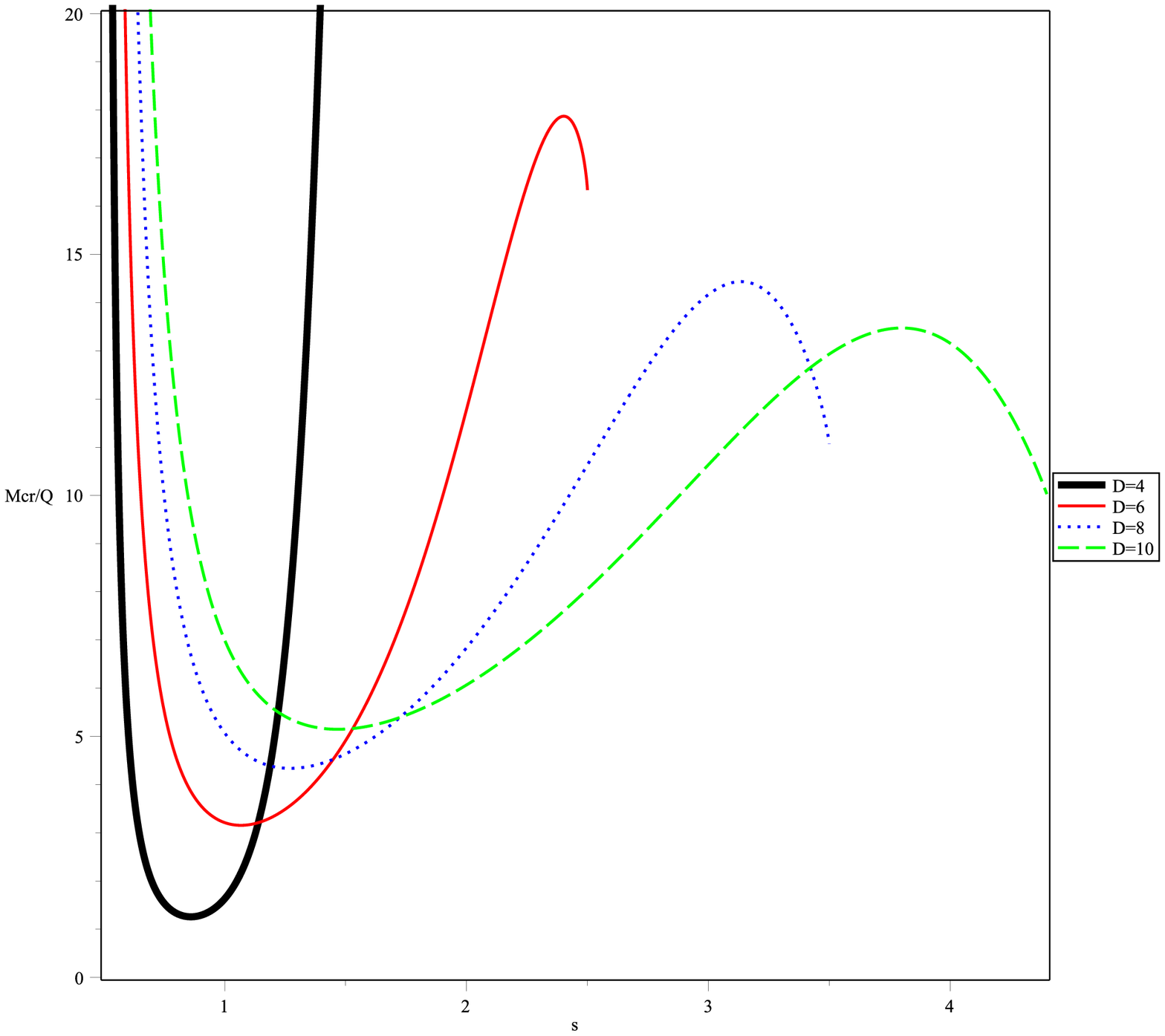} & \epsfxsize=6.5cm
\epsffile{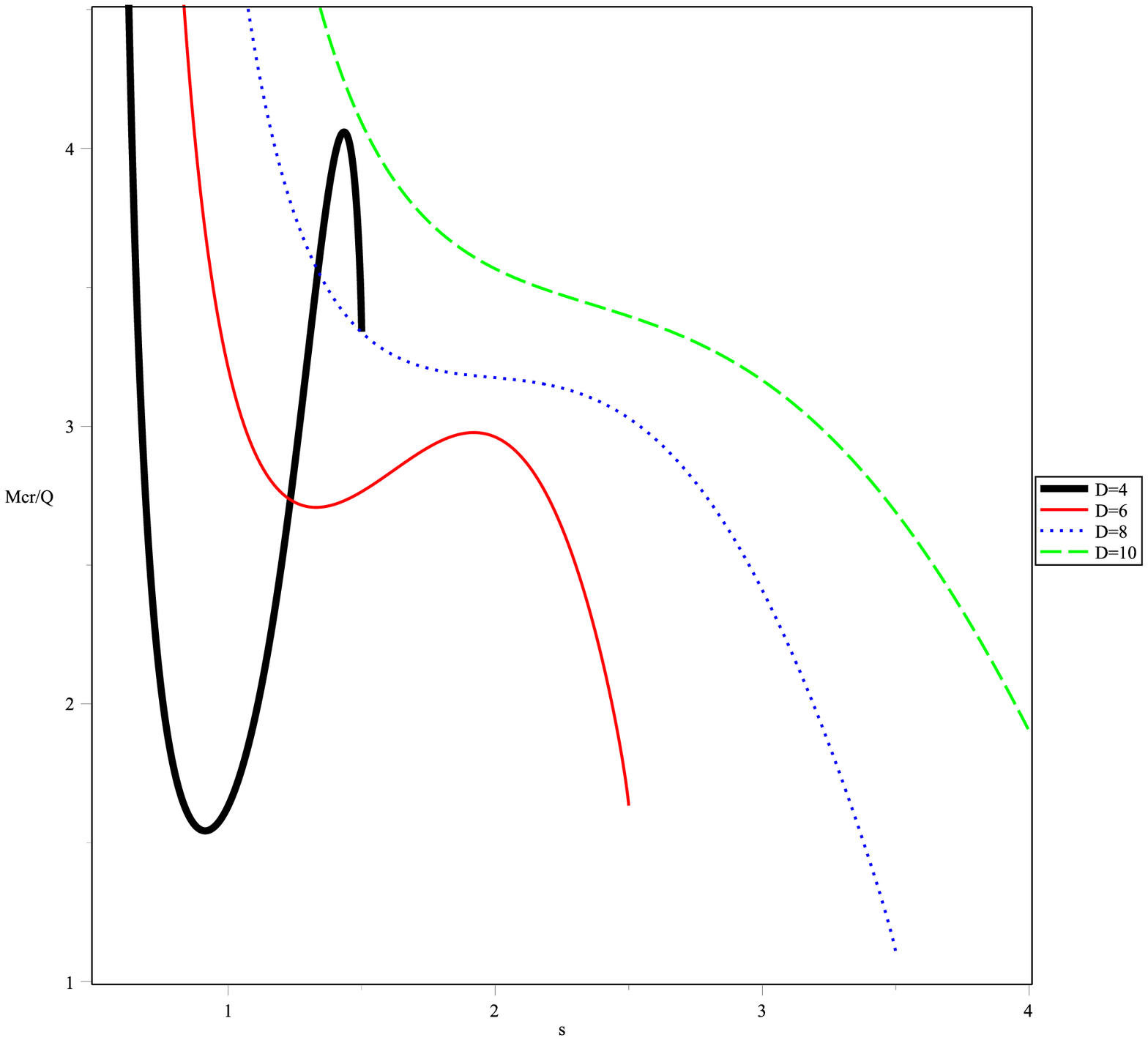}
\end{array}
$%
\caption{mass\--to\--charge ratio($\frac{M_{cr}}{Q}$) in respect
to nonlinearity parameter(s),for different dimensions. The red (solid line),
blue (dotted line), green (dash-dotted line) and black (dashed line) lines are $D=4$, $D=6$, $D=8$ and $D=10$
respectively. The right and the left figure is the case of $q=100$
and $q=10$ respectively.}
\label{fig:masstochargeratio}
\end{figure}

Finally, we look at the effective potential for different BH
charge and investigate its behavior in the limit of large $q$. The
numerical evidence of this case is represented in Table
\ref{tab22} and in the  Fig. \ref{various-q}.  The mass-to-charge
ratio of the charged BH for various nonlinearity parameter is
plotted in Fig. \ref{fig:masstochargeratio}. According to this
figure, one finds that for suitable choices of the parameters
(small charge), such ratio has an absolute minimum with a local
maximum. Interestingly, one finds that depending on the free
parameters, such ratio can be a decreasing or an increasing
function of nonlinearity parameter. This informal behavior
indicates that for a special range of $s$, changing the
nonlinearity parameter leads to various effective charge.
Comparing the identification of the electric field of a point-like
charge in the Maxwell theory and PMI one in a specific distance,
one can define an effective charge ($q_{eff}$) which has
interesting behaviors. For $r\leq1$ the effective charge is a
decreasing function of $s$ and for $s=1$ it is identified as being
the electric charge, $q$. In other words, for $1/2<s<1$ one finds
that $q_{eff}>q$, while $q_{eff}<q$ for $s>1$. An interesting
behavior appears for $r>1$. In that case $q_{eff}$ has a maximum
for special value of $s$. As the coordinate $r$ increases, the
location of maximum shifts to larger values of $s$. This maximum
starts from $s<1$ and, for sufficiently large $r$, it goes to
$s>1$. In this case ($r>1$), one obtains a nonlinearity parameter
(in addition to $s=1$), in which one gets $q_{eff}=q$ (see Fig.
3).

On the other hand, regarding large charge limit, we find that the
mass-to-charge ratio is a decreasing function of nonlinearity
parameter. This behavior indicates that, for large charge limit,
increasing the strength of nonlinearity leads to increasing the
effective charge.

%%%%%%%%%%%%%%%%%%%%%%%%%%%%%%%%%%%%%%%%%%%%%%%%%%%%%%%%%%%%%%%

%%%%%%%%%%%%%%%%%%%%%%%%%%%%%%%%%%%%%%%%%%%%%%%%%%%%%%%%%%%%%%%

%%%%%%%%%%%%%%%%%%%%%%%%%%% Qeff %%%%%%%%%%%%%%%%%%%%%%%%%%%%%%%%%%%
\begin{figure}[tbp]
$%
\begin{array}{cc}
\epsfxsize=7.5cm \epsffile{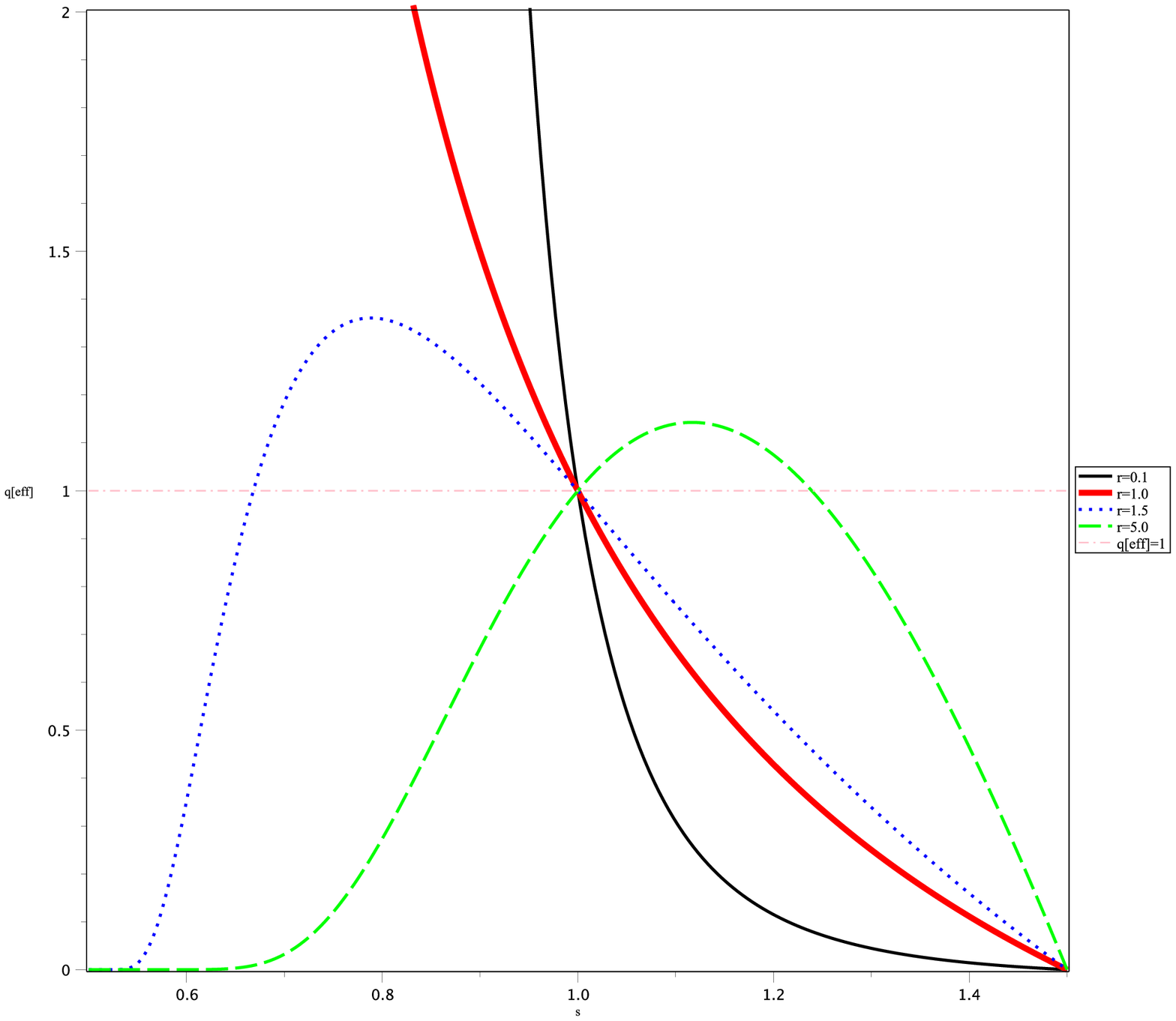} & \epsfxsize=7.5cm
\epsffile{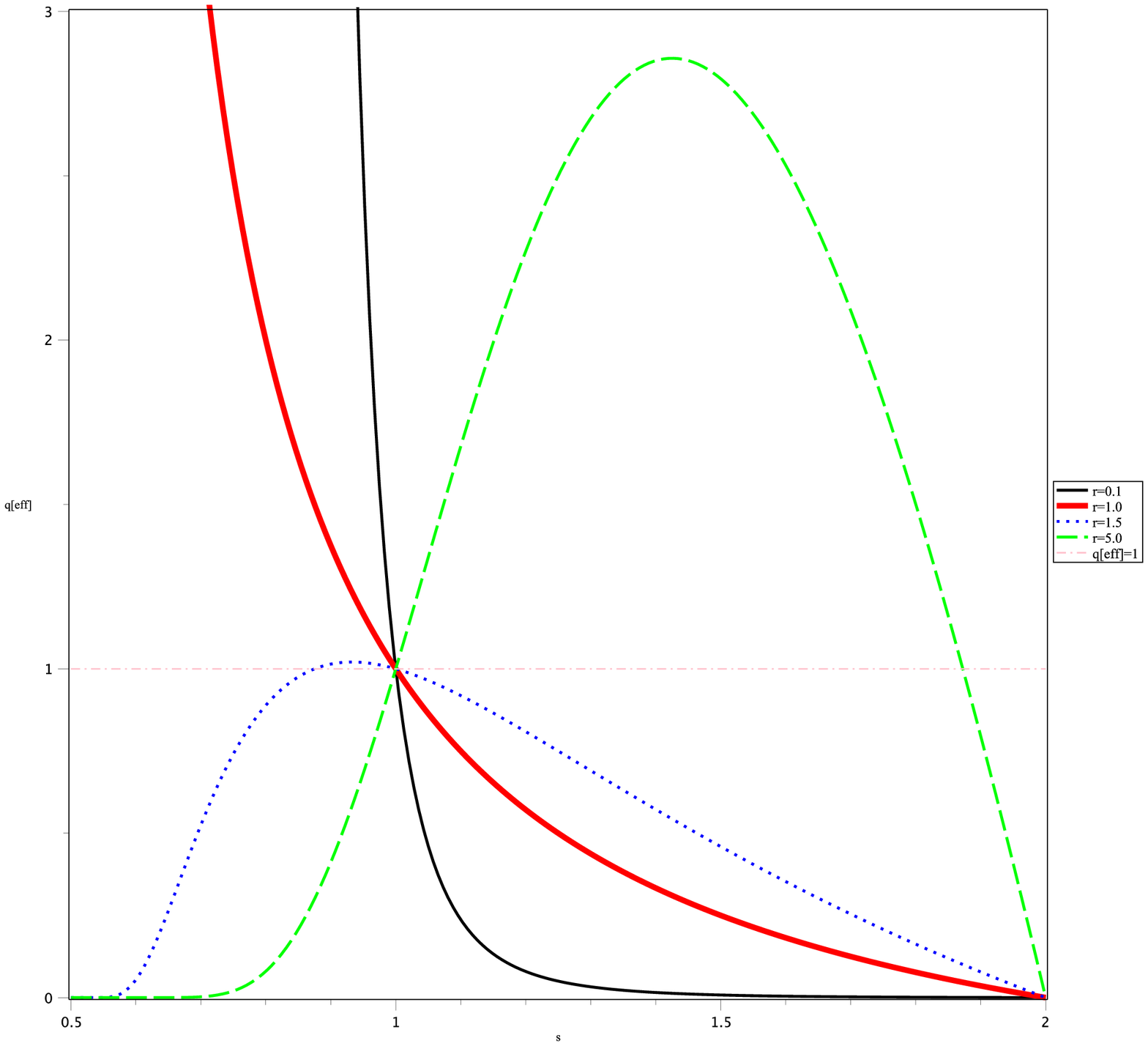}
\end{array}
$%
\caption{$q_{eff}$ versus $s$ for $q=1$, $D=4$ (left panel), $D=5$
(right panel), and $r=0.1$ (solid line), $r=1$ (bold line),
$r=1.5$ (dotted line) and $r=5$ (dashed line).} \label{Qeff}
\end{figure}
%%%%%%%%%%%%%%%%%%%%%%%%%%%%%%%%%%%%%%%%%%%%%%%%%%%%%%%%%%%%%%%%%%%%%%
Now, we are interested in studying the near horizon geometry of the
solutions by taking the large charge limit. In order to do this, one
regards the coordinate $r=r_{+}+\epsilon\sigma$ and $t=\tau/\epsilon$,
for very small $\epsilon$. Up to first order in $\epsilon$, one
can write
\begin{equation}
ds^{2}=-{(4\pi{\widetilde{T}}_{{\rm
cr}})\,\,\sigma}d\tau^{2}+\frac{1}{(4\pi{\widetilde{T}} _{{\rm
cr}})}\frac{d\sigma^{2}}{\sigma}+d{\mathbb{R}}^{D-2},\label{metric-critical}
\end{equation}
where $T=\frac{Y^{\prime}(r=r_{+})}{4\pi}$ and $r_{+}$ are
replaced by $T_{{\rm cr}}$ and $r_{{\rm cr}}$ from Eqs. (\ref{eq:
T_cr}) and (\ref{r_cr}). In addition, we considered the near
horizon limit $\epsilon\to0$ while at the same time taking the
large $q$ limit by holding ${\tilde{q}}^{s(2s-1)
/(Ds-4s+1)}=\epsilon q^{s(2s-1)/(Ds-4s+1)}$ fixed. It is notable
that ${\widetilde{T}}_{{\rm cr}}$ in the metric
($\ref{metric-critical}$) is $T_{{\rm cr}}$ in Eq. (\ref{eq:
T_cr}) being $q$ replaced by ${\tilde{q}}$ with the following
explicit form
\begin{equation}
\widetilde{T}_{cr}=\frac{4(D-3)(Ds-4s+1)(\frac{ks(D-2)^{2}(2Ds-6s+1)
\tilde{q}^{2s}}{16(D-3)(2s-1)^{2}})^{\frac{1-2s}{2(Ds-4s+1)}}}{\pi(D-2)(2Ds-6s+1)}.
\end{equation}
In addition, for $\Lambda=0$ and $A_{t}=0$, and since the
$S^{D-2}$ has infinite radius ($r_{{\rm cr}}$) at large $q$, the
metric there is essentially flat
$d{\mathbb{R}}^{D-2}=dx_{1}^{2}+\cdots+dx_{D-2}^{2}$, see
\cite{key-19,key-20}.

\section{Conclusion remarks}

In this work, we extended the results in \cite{key-20} by
analyzing the motion of a point charged particle in the framework
of NLED and AdS BHs through a probe approximation. In particular,
we considered the extended phase space by taking into account that
the cosmological constant appears as a pressure, and we studied
the effective potential. We found that such effective potential is
affected by the variation of nonlinearity parameter. We also
investigated the mass-to-charge ratio and gave an interesting
comparison between NLED and Maxwell theory with an effective
charge. Finally, we studied the near horizon geometry of the
solutions by taking the large $q$ limit.

\section{Acknowledgements }

We are indebted referees for their constructive comments. SHH and
ZST thank the Shiraz University Research Council. This work has
been supported financially by the Research Institute for Astronomy
and Astrophysics of Maragha, Iran.

\end{document}